\documentclass[11pt]{article}
\usepackage{hyperref}
\usepackage{amssymb}
\usepackage{graphicx}
\usepackage{slashed}
\usepackage{amsmath}
\usepackage{cite}
\usepackage{physics}
\usepackage{dsfont}
\usepackage{authblk}
\usepackage[section]{placeins}
\usepackage[a4paper, total={6.0in, 8in}]{geometry}
\numberwithin{equation}{section}
\title{Quantisation of Lorentz invariant scalar field theory in non-commutative space-time and its consequence}
\author{E. Harikumar \thanks{eharikumar@uohyd.ac.in}}
\author{Vishnu Rajagopal \thanks{vishnurajagopal.anayath@gmail.com}}
\affil{School of Physics, University of Hyderabad, Central University P.O, Hyderabad-500046, Telangana, India}
\date{}
\begin{document}

\maketitle
\begin{abstract}
\noindent
Quantisation of Lorentz invariant scalar field theory in Doplicher-Fredenhagen-Roberts (DFR) space-time, a Lorentz invariant, non-commutative space-time is studied. Absence of a unique Lagrangian in non-commutative space-time necessitates us to use an approach to quantisation that is based on the equations of motion alone. Using this we derive the equal time commutation relation between Doplicher-Fredenhagen-Roberts-Amorim (DFRA) scalar field and its conjugate, which has non-commutative dependent modifications, but the corresponding creation and annihilation operators obey usual algebra. We show that imposing the condition that the commutation relation between the field and its conjugate is same as that in the commutative space-time leads to a deformation of the algebra of quantised oscillators. Both these deformed commutation relations derived are valid to all orders in the non-commutative parameter. By analysing the first non-vanishing terms which are $\theta^3$ order, we show that the deformed commutaton relations scale as $1/\lambda^4$, where $\lambda$ is the length scale set by the non-commutativity of the space-time. We also derive the conserved currents for DFRA scalar field. Further, we analyse the effects of non-commutativity on Unruh effect by analysing a detector coupled to the DFRA scalar field, showing that the Unruh temperature is not modified but the thermal radiation seen by the accelerated observer gets correction due to the non-commutativity of space-time.

\end{abstract}

\section{Introduction}
Non-commutative (NC) space-time has been introduced long back in \cite{snyder}, motivated by the possibility of regularising the divergences that plague quantum field theory. NC geometry incorporates a fundamental length scale \cite{douglas} and thereby, it encodes the Planck scale physics, providing one possible way for understanding quantum gravity. Different NC space-time models have been proposed and studied thoroughly in the past decades.

Moyal space-time is one such NC space-time which is found to appear in the context of string theory \cite{seiberg}, has been studied vigorously in the recent times \cite{douglas,gross} and has a relatively quite simple commutation relation
\begin{equation}\label{moyal}
 [\hat{x}_{\mu},\hat{x}_{\nu}]=i{\theta}_{\mu\nu}.
\end{equation} 
Symmetry algebra of these NC space-times has been realised through Hopf algebra \cite{chaichian}. Such Hopf algebra structures have also been found in other NC space-times such as $\kappa$-Minkowski space-time, \cite{kappa1} where the space-time coordinates satisfy a Lie algebraic type commutation relation. Different aspects of $\kappa$-Minkowski space-time has been investigated in depth over the past few years \cite{kappa2}. 

The field theories described on NC space-time exhibit certain characteristic features such as non-local interactions and mixing of UV/IR divergences \cite{shiraz}. One major setback of the Moyal space-time is the violation of Lorentz symmetry due to the constant tensor $\theta_{\mu\nu}$. But it has been shown in \cite{doplicher1} that the Lorentz symmetry can be preserved in NC space-times by promoting this constant tensor to coordinate operator $\hat{\theta}_{\mu\nu}$ and the resulting space-time is known to be the DFR space-time. By using this formalism, it has been possible to achieve Lorentz invariance for NC field theories \cite{carlson,kase,carone,saxell}. This DFR space-time algebra has been extended further by incorporating the conjugate momenta associated with $\hat{\theta}_{\mu\nu}$ and the associated symmetry algebra is known as the DFRA algebra \cite{amorim2,amorim1,review}. In order to preserve the unitarity of the field theories on DFR space-time, it has been shown that one needs to fix $\hat{\theta}_{0i}=0$ \cite{review, gomis} and therefore, the resulting DFR space-time has $7$ dimensions.    

Different aspects of field theory models have been recently studied in DFR space-time. In \cite{amorim3} the canonical quantisation of the complex scalar fields has been studied, and its symmetries were analysed. Here a general solution for the DFRA complex scalar fields has also been constructed using the Green’s function technique. It has been shown that, in DFRA scalar model with $\phi^4$ interaction at the one-loop level, the UV/IR mixing does not exist \cite{abreu0}. The covariant Dirac equation has been derived in DFR space-time from the generalised DFRA Klein-Gordon equation \cite{amorim4}. Various other phenomenological aspects have been investigated in DFR space-time by incorporating a Lorentz invariant weight function (which depends on $\theta^2$) in the action \cite{carlson,kase,carone,saxell}. The introduction of weight function in the DFR action is necessary for the perturbative calculations \cite{carlson,kase,carone,saxell}. Since the weight function is not unique, the action of DFRA theory is also not unique, and the Lagrangian corresponding to DFRA field theories become non-unique. Such non-unique action/Lagrangian is a feature of the many NC field theories. Due to this non-uniqueness of Lagrangian, the canonical quantisation procedure may not be the appropriate method for quantising such NC field theories.

It has been shown in \cite{takahashi1,umezawa,takahashi2} that one can perform the quantisation of the field theories just from their equations of motion (EOM) alone without the knowledge of the explicit form of the Lagrangian. In this Umezawa-Takahashi scheme, one reduce every equation of motion to Klein-Gordon equation by using an operator called Klein-Gordon divisor, which helps in defining the unequal time commutation relation between the field and its adjoint, resulting in the commutation relations between the creation and annihilation operators appearing in the Fourier mode expansion of the fields. This method also helps in constructing the conserved currents, corresponding to continuous as well as discrete symmetries, directly from the equations of motion, without referring to its Lagrangian \cite{takahashi3,taka}. It also provides the possibility of calculating the conserved currents for the high spin field theories, whose Lagrangians are not defined. This quantisation method has been applied to $\kappa$-Minkowski space-time for studying the quantisation of $\kappa$-deformed scalar and $\kappa$-deformed Dirac fields \cite{vishnu1,vishnu2}.

In this work, we study the quantisation of the scalar theory in the DFR space-time using the above quantisation scheme in \cite{takahashi1,umezawa,takahashi2}. The EOM corresponding to DFRA scalar field is obtained from the action defined in \cite{amorim1}, but with Gaussian weight function. We then derive a deformed equal time commutation relation between DFRA field and its conjugate, valid to all orders in the NC parameter, by assuming the usual form for the oscillator algebra. Further, for a specific choice of the deformed DFRA oscillator algebra, valid to all orders in the NC parameter, we obtain an undeformed equal time commutation relation between DFRA field and its conjugate. By keeping only the first non-vansihing corrrections in the deformed commuation relation between DFRA field and its time derivative, we show that the deformation scales as $\frac{1}{\lambda^4}$. The deformed oscillator algebra also shares this scaling behaviour. From the EOM we also construct the energy momentum as well as angular momentum tensors corresponding to DFRA scalar field and show that the weight function introduces an asymmetry in the indices of the energy momentum tensor. The deformation of the commutation relation, which depends on the non-commutativity of the space-time captures quantum gravity effects and thus it is important to study the changes these modifications bring in. Here we analyse the implication of the deformed commutation relation between creation and annihilation operators to Unruh effect \cite{unruh,birrel,takagi,crispino}. This is studied by analysing the response of the detector coupled with the quantised DFRA scalar field. For this, we calculate the Wightman function from the vacuum expectation value of the DFRA scalar field. We show that the NC correction terms depend on the dimension of the space-time, but the Unruh temperature is unaffected.

This paper is organised into the following sections. In sec.2, we give a brief outline of the DFR space-time and the DFRA algebra associated with it. We also discuss about the symmetry of the DFR space-time and its generators. In sec.3, we start with the action corresponding to DFRA scalar field and obtain the equations of motion when a Gaussian weight function is present in action. In sec.4, we give a short summary of the quantisation procedure discussed in \cite{takahashi1,umezawa,takahashi2}. Our main result is derived in sec.5, starting from the EOM of scalar field in DFR space-time, we derive the commutation relation between this field and its conjugate, using Umezawa-Takahashi scheme. We show that this commutation relation is deformed when the corresponding quantised oscillators obey usual harmonic oscillator algebra. We also show that demanding this commutation relation to be undeformed necessitates the oscillator algebra to be deformed. In sec.6, we derive the conserved currents corresponding to translational and Lorentz symmetry for DFRA scalar field. In sec.7, we study the Unruh effect in DFR space-time by calculating the transition probability rate of the uniformly accelerating detector coupled to the DFRA scalar field. We show that the Unruh temperature is unaffected but the nature of the thermal radiation is affected by the non-commutativity. Our concluding remarks are given in sec.8.


\section{DFR space-time}

In this section, we give a brief outline of the DFR space-time, a NC space-time that respects Lorentz invariance and we summarise the underlying algebra associated with this space-time \cite{doplicher1}. 

The DFR space-time coordinates $\hat{x}_{\mu}$ and $\hat{\theta}_{\mu\nu}$, satisfy the following commutation relations
\begin{equation}\label{dfra}
\begin{split}
[\hat{x}_{\mu},\hat{x}_{\nu}]=i\hat{\theta}_{\mu\nu},&~~~[\hat{x}_{\mu},\hat{\theta}_{\nu\rho}]=0,\\
[\hat{\theta}_{\mu\lambda},\hat{\theta}_{\nu\rho}]&=0.
\end{split}
\end{equation}
All the DFR space-time coordinate operators possess canonical conjugate momenta, and we denote $\hat{p}_{\mu}$ as the canonical conjugate momenta corresponding to $\hat{x}_{\mu}$, similarly $\hat{k}_{\mu\nu}$ as the canonical conjugate momenta corresponding to $\hat{\theta}_{\mu\nu}$. The commutation relations between these coordinate operators and their conjugate momenta's are given by
\begin{equation}\label{dfra1}
\begin{split}
[\hat{x}_{\mu},\hat{p}_{\nu}]&=i\eta_{\mu\nu},~~~~[\hat{x}_{\mu},\hat{k}_{\nu\lambda}]=-\frac{i}{2}(\eta_{\mu\nu}\eta_{\rho\lambda}-\eta_{\mu\lambda}\eta_{\nu\rho})\hat{p}^{\rho},\\
[\hat{p}_{\mu},\hat{p}_{\nu}]&=0,~~~~[\hat{\theta}_{\mu\nu},\hat{k}_{\rho\lambda}]=i(\eta_{\mu\rho}\eta_{\nu\lambda}-\eta_{\mu\lambda}\eta_{\nu\rho})\\
[\hat{p}_{\mu},\hat{\theta}_{\nu\lambda}]&=0,~~~~[\hat{p}_{\mu},\hat{k}_{\nu\lambda}]=0,\\
[\hat{k}_{\mu\nu},\hat{k}_{\rho\lambda}]&=0.
\end{split}
\end{equation}
The Eq.(\ref{dfra}) and Eq.(\ref{dfra1}) form the DFRA algebra  \cite{amorim1} and they are consistent with the Jacobi identities (see \cite{amorim1}). The Lorentz generator associated with the above algebra is defined as \cite{amorim1}
\begin{equation}\label{dlorentz}
 M_{\mu\nu}=\hat{x}_{\mu}\hat{p}_{\nu}-\hat{x}_{\nu}\hat{p}_{\mu}+\frac{1}{2}\hat{\theta}_{\mu\alpha}\hat{p}^{\alpha}\hat{p}_{\nu}-\frac{1}{2}\hat{\theta}_{\nu\alpha}\hat{p}^{\alpha}\hat{p}_{\mu}-\hat{\theta}_{\mu\lambda}\hat{k}^{~\lambda}_{\nu}+\hat{\theta}_{\nu\lambda}\hat{k}^{~\lambda}_{\mu}.
\end{equation}
With these definitions given in above Eq.(\ref{dlorentz}), we obtain the following non-vanishing commutation relations \cite{amorim1} between $M_{\mu\nu}$, ${\hat p}_\lambda$ and ${\hat k}_{\alpha\beta}$, the generators of DRFA Poincare algebra,
\begin{equation}\label{dlorentz1}
\begin{split}
 [M_{\mu\nu},\hat{p}_{\lambda}]&=i(\eta_{\mu\lambda}\hat{p}_{\nu}-\eta_{\nu\lambda}\hat{p}_{\mu}),\\
[M_{\mu\nu},\hat{k}_{\alpha\beta}]&=i(\eta_{\mu\beta}\hat{k}_{\alpha\nu}-\eta_{\mu\alpha}\hat{k}_{\nu\beta}+\eta_{\nu\alpha}\hat{k}_{\beta\mu}-\eta_{\nu\beta}\hat{k}_{\alpha\mu}),\\
[M_{\mu\nu},M_{\lambda\rho}]&=i(\eta_{\mu\rho}M_{\nu\lambda}-\eta_{\nu\rho}M_{\lambda\mu}-\eta_{\mu\lambda}M_{\rho\nu}+\eta_{\nu\lambda}M_{\rho\mu}).
\end{split}
\end{equation}
From Eq.(\ref{dlorentz1}), one sees that the algebra is closed. The Casimir operator associated with the DFRA Poincare algebra is given as \cite{amorim1}
\begin{equation}\label{casimir}
 \hat{P}^2=\hat{p}_{\mu}\hat{p}^{\mu}+\lambda^2\hat{k}_{\mu\nu}\hat{k}^{\mu\nu},
\end{equation}
where $\lambda$ is the NC parameter having the dimension of length. In \cite{amorim1}, Casimir operators, like Pauli-Lubanski vector, in DFR space-time has been constructed and their properties have also been discussed in detail.


\section{DFRA scalar field theory}

In this section, we summarise the construction of action for the field theory in the DFR space-time using the definition of star product. By varying the action, we then obtain the equations of motion corresponding to the scalar field defined on the DFR space-time \footnote{Now onwards, we set $\theta_{0i}=0$ to satisfy the unitarity condition.}. Here we also recall the necessity of introducing the $\theta$-dependent weight function. 

The Moyal-star product is generalised to DFR space-time as,
\begin{equation}\label{starproduct}
 f(x,\theta)\star g(x,\theta)=e^{\frac{i}{2}\theta^{ij}\partial_{i}\partial_{j}'}f(x,\theta)g(x',\theta)\Big|_{x=x'},
\end{equation}  
where $f$ and $g$ are functions of DFR space-time coordinates. In the limit $\theta\to 0$, this reduces to usual point multiplication. Note that the non-local terms introduced by the $\star$ product depends on the derivatives with respect to the coordinate $x_{\mu}$ alone.

We define the action corresponding to the scalar field theory in DFR space-time by replacing the usual product with the star product $\star$ in Lagrangian and introducing weight function in the measure. Thus the explicit form of the non-commutative action for the DFR scalar field is given as \footnote{Here $\theta_i$ is defined as $\theta_i=\frac{1}{2}\epsilon_{ijk}\theta^{jk}$. Similarly its conjugate is defined as $k_i=\frac{1}{2}\epsilon_{ijk}k^{jk}$}
\begin{equation}\label{action1}
 S=\int d^4x~d^3\theta~W(\theta)\Big(\partial_{\mu}\phi\star\partial^{\mu}\phi+\lambda^2\partial_{\theta_i}\phi\star\partial_{\theta_i}\phi+m^2\phi\star\phi\Big).
\end{equation}
Here in the above equation, $W(\theta)$ represents the weight function, which depends on the $\theta$ coordinate alone. This weight function has been introduced in the measure in order to control the divergences associated with the perturbative calculations of the DFR field theories \cite{carlson,kase,carone,saxell}. It helps to do the calculations by allowing the $\theta$ dependent functions to have a truncated power series expansion. This weight function is taken to be an even function in $\theta$, i.e, $W(-\theta)=W(\theta)$, to have a Lorentz invariant theory \cite{carlson,kase,carone,saxell}. Gaussian function is the typical form of $W(\theta)$ and hence we use
\begin{equation}\label{weight}
 W(\theta)=\Big(\frac{1}{4\pi^2\lambda^4}\Big)^{3/2}e^{-\frac{\theta^2}{4\lambda^4}},
\end{equation}
in this study. In above expression, $(\frac{1}{4\pi^2\lambda^4})^{3/2}$ represents the normalisation factor. Other choices for the weight function satisfying the above conditions are considered in \cite{carone} showing the weight function is not unique.   

From Eq.(\ref{starproduct}), it is easy to see that the Moyal product satisfy 
\begin{equation}\label{spproduct}
 \int d^4x~d^3\theta~W(\theta)~f(x,\theta)\star g(x,\theta)=\int d^4x~d^3\theta~W(\theta)~f(x,\theta)g(x,\theta).
\end{equation}
After using the above property in Eq.(\ref{action1}), we get the action for DFRA scalar field theory as
\begin{equation}\label{action2}
 S=\int d^4x~d^3\theta~W(\theta)\Big(\partial_{\mu}\phi\partial^{\mu}\phi+\lambda^2\partial_{\theta_i}\phi\partial_{\theta_i}\phi+m^2\phi^2\Big).
\end{equation}
Here we observe that the Lagrangian contains a $\lambda^2$ dependent term signifying the $\theta$ contribution of the DFRA action. We get the commutative action corresponding to the scalar field in the limit $\lambda\to 0$ and using the result $\lim_{\lambda\to 0}\frac{e^{-\frac{\theta_i^2}{4\lambda^4}}}{\lambda^2}=\delta(\theta_i)$ \cite{abreu1}.

The equations of motion corresponding to the scalar field in DFR space-time, by varying the action given in Eq.(\ref{action2}) is
\begin{equation}\label{eom1}
 \Big(\Box+\lambda^2\Box_{\theta}-m^2\Big)\phi(x,\theta)+\lambda^2\partial_{\theta_i}\Big(\ln W(\theta)\Big)\partial_{\theta_i}\phi(x,\theta)=0.
\end{equation} 
From Eq.(\ref{eom1}), we see that the last term on the LHS depends on the derivative of the weight function and therefore this term vanishes when weight function becomes unity, i.e., $W(\theta)=1$ as in \cite{amorim3}. Thus the EOM given in Eq.(\ref{eom1}) reduces to the EOM obtained from the quadratic Casimir, i.e, Eq.(\ref{casimir}), when $W(\theta)=1$ \cite{amorim3}. We obtain the commutative scalar field EOM from Eq.(\ref{eom1}) in the $\lim \lambda\to 0$.

After substituting the explicit form of the weight function given in Eq.(\ref{weight}), in the above, we get
\begin{equation}\label{eom2}
 \Big(\Box+\lambda^2\Box_{\theta}-m^2\Big)\phi(x,\theta)-\frac{\theta_i}{2\lambda^2}\partial_{\theta_i}\phi(x,\theta)=0.
\end{equation} 
The above equation reduces to the commutative scalar field EOM in the $\lim\lambda\to 0$ with $\theta=0$. It is to be noted that the first three terms on the LHS of the above equation correspond to the EOM obtained from the Casimir of the DFRA algebra \cite{amorim1,abreu1}. The dynamics and the canonical quantisation of the DFRA scalar field obeying that equation of motion has been studied in detail in \cite{amorim3}, starting from Lagrangian.

We note that the equations of motion obtained from the DFRA scalar field action given above and one obtained from the quadratic Casimir of the symmetry algebra are not the same. This difference is due to the presence of the weight function in action. Hence the equations of motion obtained from the action depends on the choice of the weight function (which is not unique). This leads to the non-uniqueness of equations of motion for the scalar field in DFR space-time (and all of these equations of motion reduces to the correct commutative limit). Thus it is appropriate to study the quantisation of the DFRA field theories using their equations of motion alone, rather than using the canonical scheme.

\section{Quantisation using EOM} 
In this section, we briefly outline the quantisation procedure discussed in \cite{takahashi1,umezawa,takahashi2} which leads to the derivation of the (anti-) commutation relations between field and its adjoint using the EOM satisfied by the field. Note that one does not use the canonical conjugate momenta as this approach does not rely on Lagrangian. This quantisation scheme requires only EOM for quantising the field theories. Further, this method also helps in obtaining the conserved currents associated with the symmetry transformations, including discrete ones. The quantisation of high spin field theories, which are usually defined through the EOM, was studied using this method \cite{takahashi1,umezawa,takahashi2}.

The equations of motion satisfied by a field $\phi(x)$ is represented using an operator $\Lambda(\partial)$ as
\begin{equation}\label{lambda}
 \Lambda(\partial)\phi(x)=0.
\end{equation}
Here $\Lambda(\partial)$ is a polynomial in $\partial_{\mu}$ operators and in general the $\Lambda(\partial)$ operator is defined as \cite{takahashi1,takahashi2}
\begin{equation}\label{lambda1}
 \begin{split}
\Lambda(\partial)&=\sum_{l=0}^{N}\Lambda_{\mu_1\mu_2....\mu_l}\partial^{\mu_1}\partial^{\mu_2}....\partial^{\mu_l}\\
&=\Lambda+\Lambda_{\mu}\partial^{\mu}+\Lambda_{\mu\nu}\partial^{\mu}\partial^{\nu}+\Lambda_{\mu\nu\rho}\partial^{\mu}\partial^{\nu}\partial^{\rho}+..........+\Lambda_{\mu_1\mu_2\mu_3....\mu_N}\partial^{\mu_1}\partial^{\mu_2}\partial^{\mu_3}....\partial^{\mu_N}.
\end{split}
\end{equation} 
According to this procedure, every free field equations of motion, given as in Eq.(\ref{lambda}), can be reduced to usual Klein-Gordon equation by acting it with an operator called Klein-Gordan divisor, $d(\partial)$. Thus we have 
\begin{equation}\label{kg1}
 d(\partial)\Lambda(\partial)=\Box-m^2.
\end{equation}
This Klein-Gordon divisor should commute with the EOM, i.e, $[\Lambda(\partial),d(\partial)]=0$ and it should have non-zero eigen values \cite{takahashi1,takahashi2}, so that $d(\partial)$ can be inverted. For a usual Klein-Gordon field, in the commutative space-time, $d(\partial)$ is the identity operator i.e, $d(\partial)=\mathds{I}$. The details of the construction of $d(\partial)$ for generic field theories are given in \cite{takahashi2}.  

The field operator satisfying the EOM, given in Eq.(\ref{lambda}), is decomposed into Fourier modes, using the creation and annihilation operators, i.e, $a^{\dagger}(p),~a(p)$ as 
\begin{equation}\label{decomp}
 \phi(x)=\int\frac{d^3p}{\sqrt{(2\pi)^32E_p}}\Big(u_p(x)a(p)+u^*_p(x)a^{\dagger}(p)\Big).
\end{equation}
Here $u_p(x)$ satisfies the EOM defined in Eq.(\ref{lambda}), i.e, $\Lambda(\partial)u_p(x)=0$. These creation and annihilation operators assumed to satisfy the following commutation relations, 
\begin{equation}\label{alg}
 [a(p),a(p')]=[a^{\dagger}(p),a^{\dagger}(p')]=0,~~ [a(p),a^{\dagger}(p')]=\delta^3(p-p').
\end{equation}
Now in this quantisation procedure, one writes down the (unequal time) commutation relation between the field and its adjoint operator, using the Klein-Gordon divisor $d(\partial)$ as
\begin{equation}\label{comm1}
[\phi(x),\bar{\phi}(x')]=id(\partial)\Delta(x-x'),
\end{equation}
where 
\begin{equation*}\label{delta0}
 \Delta(x-x')=\int\frac{d^3p}{(2\pi)^32E_p}\big(e^{-ip(x-x')}-e^{ip(x-x')}\big).
\end{equation*}
Here in the Eq.(\ref{comm1}), if $\phi(x)$ is a fermionic field then the above commutation relations are replaced with the corresponding anti-commutation relations as per the spin-statistics theorem.

Note that one can get an equal-time commutation relation between field, $\phi(x)$ and its time-derivative, $\partial_t\phi(x)$. This is done by applying the time derivative, i.e, $\partial_{t'}$, on Eq.(\ref{comm1}) and then setting, $t=t'$. We will use this method for obtaining the equal-time commutation relation between DFRA scalar field and its derivative in the subsequent sections.
 
Another important feature of this quantisation scheme is that it provides a constructive method to derive the conserved charges associated with symmetry transformations, just from the EOM alone (unlike Noether's method, here one does construct conserved current and charges corresponding to discrete symmetries also). For this purpose one define an operator $\Gamma_{\mu}(\partial,-\overleftarrow{\partial})$ as
\begin{equation}\label{gamma}
\begin{split}
 \Gamma_{\mu}(\partial,-\overleftarrow{\partial})&=\sum_{l=0}^{N-1}\sum_{i=0}^{l}\Lambda_{\mu\mu_1.....\mu_l}\partial_{\mu_1}.....\partial_{\mu_i}(-\overleftarrow{\partial}_{\mu_{i+1}})......(-\overleftarrow{\partial}_{\mu_l})\\
&=\Lambda_{\mu}+\Lambda_{\mu\nu}(\partial^{\nu}-\overleftarrow{\partial}^{\nu})+\Lambda_{\mu\nu\rho}(\partial^{\nu}\partial^{\rho}-\partial^{\nu}\overleftarrow{\partial}^{\rho}+\overleftarrow{\partial}^{\nu}\overleftarrow{\partial}^{\rho})+......
\end{split}
\end{equation}
and using Eq.(\ref{lambda1}) it can be shown that $\Gamma_{\mu}(\partial,-\overleftarrow{\partial})$ satisfies an identity \cite{takahashi1,takahashi2} given by
\begin{equation}\label{iden}
 (\partial^{\mu}+\overleftarrow{\partial}^{\mu})\Gamma_{\mu}(\partial,-\overleftarrow{\partial})=\Lambda(\partial)-\Lambda(-\overleftarrow{\partial}).
\end{equation}
Using this $\Gamma_{\mu}(\partial,-\overleftarrow{\partial})$ operator, the conserved current $J_{\mu}$ associated with a symmetry transformation is defined as \cite{takahashi1,takahashi2}
\begin{equation}\label{current}
 J_{\mu}=\bar{\phi}(x)\Gamma_{\mu}(\partial,-\overleftarrow{\partial})\delta\phi(x),
\end{equation}
where $\delta\phi(x)$ refers to the variation in the field under the symmetry transformation. Using Eq.(\ref{current}) we can construct the energy-momentum tensor as \cite{takahashi1}
\begin{equation}\label{taka-em}
 T_{\mu\nu}=\bar{\phi}(x)\Gamma_{\mu}(\partial,-\overleftarrow{\partial})\partial_{\nu}\phi(x)
\end{equation}
and generators of Lorentz transformation as
\begin{equation}\label{taka-ang}
 M_{\mu\nu}=\int d^3x~\bar{\phi}(x)\Gamma_{0}(\partial,-\overleftarrow{\partial})x_{\mu}\partial_{\nu}\phi(x),
\end{equation}
respectively. 

This method has also been used to construct the conserved currents corresponding to the discrete symmetries \cite{takahashi2}. The conserved current associated with the discrete symmetries of the $\kappa$-deformed Dirac field, a NC field theory, has been derived in \cite{vishnu2} using this procedure, showing that the charge conjugation is not a symmetry of $\kappa$-Dirac field \cite{hari}.

\section{Quantisation of DFRA scalar field}

From the discussions in sec.3, we have seen that the action/Lagrangian of the scalar field in DFR space-time is not unique. Thus it is natural to study the quantisation of DFR field theories using their EOM, rather than the action. In this section we use the approach of \cite{takahashi1,umezawa,takahashi2} to quantise the scalar field in DFR space-time by using its equations of motion given in Eq.(\ref{eom2}) as the starting point. We also derive the conserved currents associated with the scalar field in DFR space-time. 

We first generalise the definition for $\Lambda(\partial)$ operator to DFR space-time as $\Lambda(\partial,\partial_{\theta})$. Therefore using Eq.(\ref{lambda1}) we get the expression for $\Lambda(\partial,\partial_{\theta})$ operator as
\begin{equation}\label{dfr-lambda2}
 \begin{split}
\Lambda(\partial,\partial_{\theta})&=\sum_{l=0}^{N}\Lambda_{A_1A_2....A_l}\partial^{A_1}\partial^{A_2}....\partial^{A_l}\\
&=\Lambda_0+\Lambda_{A}\partial^{A}+\Lambda_{AB}\partial^{A}\partial^{B}+\Lambda_{ABC}\partial^{A}\partial^{B}\partial^{C}+..........+\Lambda_{A_1A_2A_3....A_N}\partial^{A_1}\partial^{A_2}\partial^{A_3}....\partial^{A_N},
\end{split}
\end{equation} 
where $A=(\mu,\theta_i)$ and $\partial_A=(\partial_{\mu},\lambda\partial_{\theta_i})$.

Now we adapt the Eq.(\ref{kg1}) for the DFR space-time by replacing $\Lambda(\partial)$ as well as $d(\partial)$ with $\Lambda(\partial,\partial_{\theta})$ and $d(\partial,\partial_{\theta})$, respectively. Correspondingly, we replace the Klein-Gordon equation on RHS of Eq.(\ref{kg1}) with Eq.(\ref{eom2}). Thus we get
\begin{equation}\label{dfr-lambda}
 d(\partial,\partial_{\theta})\Lambda(\partial,\partial_{\theta})=\Box+\lambda^2\Box_{\theta}-m^2-\frac{\theta}{2\lambda^2}\partial_{\theta}.
\end{equation}
Now for the scalar field in DFR space-time, satistfying the EOM given in Eq.(\ref{eom2}), we have 
\begin{equation}\label{dfr-lambda1}
 \Lambda(\partial,\partial_{\theta})=\Box+\lambda^2\Box_{\theta}-m^2-\frac{\theta}{2\lambda^2}\partial_{\theta} \textnormal{ and }d(\partial,\partial_{\theta})=\mathds{I}.
\end{equation}
By comparing Eq.(\ref{dfr-lambda1}) with Eq.(\ref{dfr-lambda2}), we get the non-vanishing components of $\Lambda(\partial,\partial_{\theta})$ as 
\begin{equation}\label{dfr-lambda3}
 \Lambda_0=-m^2,~\Lambda_A=\Big(0,-\frac{\epsilon\theta_i}{2\lambda^3}\Big),~\Lambda_{AB}=diag(-1,1,1,1,1,1,1).
\end{equation}
In order to decompose the DFRA scalar field, $\phi(x,\theta)$ into creation and annihilation operators, we need to find the explicit solution of Eq.(\ref{eom2}), i.e, $u(x,\theta)$ satisfying $\Lambda(\partial,\partial_{\theta})u(x,\theta)=0$. For deriving this solution we start with the ansatz for $u(x,\theta)$ 
\begin{equation}\label{ansatz}
 u(x,\theta)=e^{-ipx}f(\theta).
\end{equation}
With $f(\theta)\Big|_{\theta=0}=1$, we find that $u(x,\theta)\Big|_{\theta=0}=u(x)$ which obeys the commutative field equation $\Lambda(\partial)u(x)=0$.

Substituting this in the equation $\Lambda(\partial,\partial_{\theta})u(x,\theta)=0$, we get an equation for $f(\theta)$ as
\begin{equation}\label{f}
 \Big(\partial^2_{\theta}-\frac{\theta}{2\lambda^4}\partial_{\theta}-p^2-m^2\Big)f(\theta)=0.
\end{equation}
We solve this using the power series, i.e.,
\begin{equation}\label{f1}
 f(\theta)=\sum_{n=0}^{\infty}a_n\theta^n.
\end{equation}
Here we note that $a_0=1$ for $f(\theta)\Big|_{\theta=0}=1$ and this allows us to fix $a_0=1$.

Using Eq.(\ref{f1}) in Eq.(\ref{f}) and using the DFR dispersion relation (see Eq.(\ref{eom2})), i.e, $p^2+m^2+\lambda^2 k^2+\frac{ik\cdot \theta}{2\lambda^2}=0$\footnote{Note we use $k\theta$ for $k\cdot\theta$ from now onwards.}, we get the solution for $f(\theta)$ as
\begin{equation}\label{f2}
 f(\theta)=a_0\Big(\cos k\theta+\sum_{n=1}^{\infty}A_n(k)(k\theta)^{2n}\Big)+a_1\Big(\sin k\theta+\sum_{n=1}^{\infty}B_n(k)(k\theta)^{2n+1}\Big),
\end{equation}
where
\begin{equation}\label{A}
\begin{split}
 A_n(\lambda,k,\theta)&=\frac{(-1)^n}{(2n)!}\bigg[\Pi_{j=0}^{n-1}\Big(1-\frac{ j}{\lambda^4 k^2}+\frac{ik\theta}{2\lambda^2}\Big)-1\bigg],\\
 B_n(\lambda,k,\theta)&=\frac{(-1)^n}{(2n+1)!}\bigg[\Pi_{j=0}^{n-1}\Big(1-\frac{ (2j+1)}{2\lambda^4 k^2}+\frac{ik\theta}{2\lambda^2}\Big)-1\bigg].
\end{split}
\end{equation}
Note $(k\theta)$ and $k\lambda^2$ are dimensionless. Here $A_n$ and $B_n$ are coming only from the weight function dependent terms in Eq.(\ref{f}). Therefore $A_n$ and $B_n$ vanishes when this weight function becomes unity. In the $\lim\lambda\to 0$ with $\theta=0$, $A_n$ and $B_n$ vanishes.

One can show that this solution is invariant under the Lorentz transformation using the transformation rules given in \cite{amorim1}. Hence we can decompose the field operator $\phi(x,\tilde{\theta})$ \footnote{We define $\tilde{\theta}=\theta/\lambda$ and $\tilde{k}=\lambda k$. In general we denote DFRA scalar field as $\phi(x,{\theta})$. But when we decompose the field into momentum space we use $\phi(x,\tilde{\theta})$} into positive and negative frequency solutions with the help of creation and annihilation operator as in the commutative space-time. Now we generalise the decompostion of scalar field, $\phi(x,\tilde{\theta})$, in DFR space-time, using the above expressions, i.e, Eq.(\ref{ansatz}) and Eq.(\ref{f2}),
\begin{equation}\label{phi}
\begin{split}
 \phi(x,\tilde{\theta})=&\int \frac{d^3p}{\sqrt{(2\pi)^3}}\frac{d^3\tilde{k}}{\sqrt{(2\pi)^3}}\frac{1}{\sqrt{2\omega(p,\tilde{k})}}\Bigg[e^{-ipx}\bigg(a_0\Big(\cos (\tilde{k}\tilde{\theta})+\sum_{n=1}^{\infty}A_n(\lambda,\tilde{k},\tilde{\theta})(\tilde{k}\tilde{\theta})^{2n}\Big)+\\
&a_1\Big(\sin (\tilde{k}\tilde{\theta})+\sum_{n=1}^{\infty}B_n(\lambda,\tilde{k},\tilde{\theta})(\tilde{k}\tilde{\theta})^{2n+1}\Big)\bigg)a(p,\tilde{k})+h.c\Bigg]
\end{split}
\end{equation}
In the above equation we re-express the cosine and sine functions using Euler's identity and after re-arranging we get
\begin{equation}\label{phi-1}
\begin{split}
 \phi(x,\tilde{\theta})=&\int \frac{d^3p}{\sqrt{(2\pi)^3}}\frac{d^3\tilde{k}}{\sqrt{(2\pi)^3}}\frac{1}{\sqrt{2\omega(p,\tilde{k})}}\Bigg[e^{-ipx}\bigg(\frac{1}{2}(a_0+ia_1)e^{-ik\theta}+\frac{1}{2}(a_0-ia_1)e^{ik\theta}+\\
&a_0\sum_{n=1}^{\infty}A_n(\lambda,\tilde{k},\tilde{\theta})(\tilde{k}\tilde{\theta})^{2n}+a_1\sum_{n=1}^{\infty}B_n(\lambda,\tilde{k},\tilde{\theta})(\tilde{k}\tilde{\theta})^{2n+1}\bigg)a(p,\tilde{k})+h.c\Bigg]
\end{split}
\end{equation}
Now in the second term we change $\tilde{k}\to -\tilde{k}$ and perform the integration to get a plane wave mode for the DFRA scalar field as
\begin{equation}\label{phi0}
\begin{split}
 \phi(x,\tilde{\theta})=&\int \frac{d^3p}{\sqrt{(2\pi)^3}}\frac{d^3\tilde{k}}{\sqrt{(2\pi)^3}}\frac{1}{\sqrt{2\omega(p,\tilde{k})}}\Bigg(e^{-ipx}\bigg(a_0e^{-i\tilde{k}\tilde{\theta}}+
a_0\sum_{n=1}^{\infty}A_n(\lambda,\tilde{k},\tilde{\theta})(\tilde{k}\tilde{\theta})^{2n}\\
&+a_1\sum_{n=1}^{\infty}B_n(\lambda,\tilde{k},\tilde{\theta})(\tilde{k}\tilde{\theta})^{2n+1}\bigg)a(p,\tilde{k})+h.c\Bigg).
\end{split}
\end{equation}
Here we see that when the weight function reduces to unity, the above equation, Eq.(\ref{phi0}), becomes the plane wave solution in $d=7$ as shown in \cite{amorim3,abreu1} for $a_0=1$ and for arbitrary $a_1$. Taking this as a criterion, set $a_0=1$ (see the discussion after Eq.(\ref{f1})) and express the DFRA scalar field as 
\begin{equation}\label{phi1}
\begin{split}
 \phi(x,\tilde{\theta})=&\int \frac{d^3p}{\sqrt{(2\pi)^3}}\frac{d^3\tilde{k}}{\sqrt{(2\pi)^3}}\frac{1}{\sqrt{2\omega(p,\tilde{k})}}\Bigg(\bigg(e^{-i(px+\tilde{k}\tilde{\theta})}+
e^{-ipx}\mathcal{G}(\tilde{k},\tilde{\theta})\bigg)a(p,\tilde{k})+h.c\Bigg),
\end{split}
\end{equation}
where
\begin{equation}\label{gk}
 \mathcal{G}(\tilde{k},\tilde{\theta})=\sum_{n=1}^{\infty}\Big(A_n(\lambda,\tilde{k},\tilde{\theta})(\tilde{k}\tilde{\theta})^{2n}+a_1B_n(\lambda,\tilde{k},\tilde{\theta})(\tilde{k}\tilde{\theta})^{2n+1}\Big),
\end{equation}  
and it vanishes when the weight function becomes unity as well as in the $\lim\lambda\to 0$ with ${\theta}=0$. Therefore we see that in the limit $\lambda\to 0$ the DFRA scalar field reduces to the commutative scalar field.

The commutation relation between a DFRA field and its adjoint can be obtained by generalising the expression given in Eq.(\ref{comm1}) to DFR space-time. This is done by replacing $\Delta(x-x')$ with $\Delta(x,x';\theta,\theta')$ as well as $d(\partial)$ with $d(\partial,\partial_{\theta})$. Thus we get
\begin{equation}\label{dfr-comm}
 \Big[\phi(x,\tilde{\theta}),\bar{\phi}(x',\tilde{\theta}')\Big]=id(\partial,\partial_{\theta})\Delta(x,x';\tilde{\theta},\tilde{\theta}').
\end{equation}
For the DFRA scalar field we have seen that $d(\partial,\partial_{\theta})=\mathds{I}$ and therefore Eq.(\ref{dfr-comm}) becomes
\begin{equation}\label{dfr-comm1}
 \Big[\phi(x,\tilde{\theta}),\bar{\phi}(x',\tilde{\theta}')\Big]=i\Delta(x,x';\tilde{\theta},\tilde{\theta}').
\end{equation}  
Note that in the above expression, $\Delta(x,x';\tilde{\theta},\tilde{\theta}')$ is an unknown function and this can be derived by calculating the LHS of Eq.(\ref{dfr-comm1}) explicitly. So now we evaluate the LHS of Eq.(\ref{dfr-comm1}) by substituting Eq.(\ref{phi1}) and we get
\begin{equation}\label{dfr-LHS}
\begin{split}
 \Big[\phi(x,\tilde{\theta}),{\phi}(x',\tilde{\theta}')\Big]=&\int\frac{d^3p}{\sqrt{(2\pi)^3}}\frac{d^3\tilde{k}}{\sqrt{(2\pi)^3}}\frac{d^3p'}{\sqrt{(2\pi)^3}}\frac{d^3\tilde{k}'}{\sqrt{(2\pi)^3}}\frac{1}{\sqrt{2\omega(p,\tilde{k})}}\frac{1}{\sqrt{2\omega(p',\tilde{k}')}}\\
&\bigg(u_{p,\tilde{k}}(x,\tilde{\theta})u_{p',\tilde{k}'}^*(x',\tilde{\theta}')[a(p,\tilde{k}),a^{\dagger}(p',\tilde{k}')]-u_{p,\tilde{k}}^*(x,\tilde{\theta})u_{p',\tilde{k}'}(x,\tilde{\theta}')[a(p',\tilde{k}'),a^{\dagger}(p,\tilde{k})]\bigg).
\end{split}
\end{equation}     
We assume that the creation and annihilation operators for the scalar field in DFR space-time satisfy the following commutation relation
\begin{equation}\label{dfr-undef}
 [a(p,\tilde{k}),a^{\dagger}(p',\tilde{k}')]=\delta^3(p-p')\delta^3(\tilde{k}-\tilde{k}').
\end{equation}
By substituting the above undeformed oscillator algebra in Eq.(\ref{dfr-LHS}), we get the unequal time commutator between DFRA scalar field and its adjoint as
\begin{equation}\label{dfr-LHS1}
\begin{split}
 \Big[\phi(x,\tilde{\theta}),{\phi}(x',\tilde{\theta}')\Big]=&\int\frac{d^3p}{{(2\pi)^3}}\frac{d^3\tilde{k}}{{(2\pi)^3}}\frac{1}{{2\omega(p,\tilde{k})}}
\bigg(e^{-ip(x-x')}e^{-i\tilde{k}(\tilde{\theta}-\tilde{\theta}')}-e^{ip(x-x')}e^{i\tilde{k}(\tilde{\theta}-\tilde{\theta}')}+\\
&e^{-ip(x-x')}\Big(e^{-i\tilde{k}\tilde{\theta}}\mathcal{G}^*(\tilde{k},\tilde{\theta}')+e^{i\tilde{k}\tilde{\theta}'}\mathcal{G}(\tilde{k},\tilde{\theta})+\mathcal{G}(\tilde{k},\tilde{\theta})\mathcal{G}^*(\tilde{k},\tilde{\theta}')\Big)-\\
&e^{ip(x-x')}\Big(e^{i\tilde{k}\tilde{\theta}}\mathcal{G}(\tilde{k},\tilde{\theta}')+e^{-i\tilde{k}\tilde{\theta}'}\mathcal{G}^*(\tilde{k},\tilde{\theta})+\mathcal{G}^*(\tilde{k},\tilde{\theta})\mathcal{G}(\tilde{k},\tilde{\theta}')\Big) \bigg).
\end{split}
\end{equation}
By comparing Eq.(\ref{dfr-LHS1}) with Eq.(\ref{dfr-comm1}), we get the explicit form of $i\Delta(x,x';\tilde{\theta},\tilde{\theta}')$. 
Note that the last six terms on the RHS of Eq.(\ref{dfr-LHS1}) are due to the weight function present in the action.

We now take the time derivative of the field operator and then evaluate its commutator with $\phi(x,\tilde{\theta})$ at equal time. Thus we get
\begin{equation}\label{dfr-LHS2}
\begin{split}
 \Big[\phi(x,\tilde{\theta}),\frac{d\phi(x',\tilde{\theta}')}{dt'}\Big]\bigg|_{t=t'}=&i\delta^3(x-x')\delta^3(\tilde{\theta}-\tilde{\theta}')+
i\delta^3(x-x')\int\frac{d^3\tilde{k}}{(2\pi)^3}\bigg(\frac{e^{-i\tilde{k}\tilde{\theta}}\mathcal{G}^*(\tilde{k},\tilde{\theta}')}{2}+\\
&\frac{e^{i\tilde{k}\tilde{\theta}'}\mathcal{G}(\tilde{k},\tilde{\theta})}{2}+\frac{e^{i\tilde{k}\tilde{\theta}}\mathcal{G}(\tilde{k},\tilde{\theta}')}{2}+\frac{e^{-i\tilde{k}\tilde{\theta}'}\mathcal{G}^*(\tilde{k},\tilde{\theta})}{2}+\frac{\mathcal{G}(\tilde{k},\tilde{\theta})\mathcal{G}^*(\tilde{k},\tilde{\theta}')}{2}+\\
&\frac{\mathcal{G}(\tilde{k},\tilde{\theta}')\mathcal{G}^*(\tilde{k},\tilde{\theta})}{2}\bigg).
\end{split}
\end{equation} 
Note that all the terms on RHS of the above equation depend on $x_{\mu}$ and $\theta$. This gives the deformed (equal-time) commutation relation between the DFRA field and its (time) derivative, valid to all orders in $\theta/\lambda$. We call this commutation relation to be deformed because of the presence of the weight function dependent term $\mathcal{G}(\tilde{k},\tilde{\theta})$ in it. Here it is noted that the form of the last six terms changes with the choice of weight function. This result is in agreement with that obtained in \cite{amorim2}, when the weight function is set to unity (the RHS of the above equation still depends on NC coordinate $\theta$ through the first term on RHS). 

To show the modification due to the non-commutativity of space-time more clrealy, we now write down the deformed commutation relation, Eq.(\ref{dfr-LHS2}), keeping only first non-vanishing terms in $\tilde{\theta}$. Thus we find
\begin{equation}\label{dfr-LHS2a}
\begin{split}
 \Big[\phi(x,\tilde{\theta}),\frac{d\phi(x',\tilde{\theta}')}{dt'}\Big]\bigg|_{t=t'}=&i\delta^3(x-x')\delta^3(\tilde{\theta}-\tilde{\theta}')+
i\delta^3(x-x')\int\frac{d^3\tilde{k}}{(2\pi)^3}\bigg(\frac{(\tilde{k}\tilde{\theta'})^3}{12\tilde{k}^2\lambda^2}+\frac{(\tilde{k}\tilde{\theta})^3}{12\tilde{k}^2\lambda^2}\bigg).
\end{split}
\end{equation}
We observe that the first non-vanishing correction term in the deformation (the second term in the above) of the above commutation relation depends on $\theta^3$ terms. Note that this correction term also depends on  non-commutative length scale as $1/\lambda^4$.
  
The DFRA oscillator algebra defined in Eq.(\ref{dfr-undef}) is undeformed. Next let us assume that the DFRA oscillator algebra is deformed while the commutation relation between the field and its adjoint is not deformed. Thus we take the following form for the deformed oscillator algebra
\begin{equation}\label{dfr-def}
 [a(p,\tilde{k}),a^{\dagger}(p',\tilde{k}')]=\delta^3(p-p')\delta^3(\tilde{k}-\tilde{k}')g(\tilde{k}),
\end{equation}
where $g(\tilde{k})$ is an unknown function whose explicit form has to be determined.

Let us substitute this deformed oscillator algebra in Eq.(\ref{dfr-LHS}) and repeat the above procedure, and derive the commutation relation between DFRA scalar field and its adjoint as
\begin{equation}\label{dfr-LHS4}
\begin{split}
 \Big[\phi(x,\tilde{\theta}),\phi(x',\tilde{\theta}')\Big]=&\int\frac{d^3p}{{(2\pi)^3}}\frac{d^3\tilde{k}}{{(2\pi)^3}}\frac{1}{{2\omega(p,\tilde{k})}}g(\tilde{k})
\bigg(e^{-ip(x-x')}e^{-i\tilde{k}(\tilde{\theta}-\tilde{\theta}')}-e^{ip(x-x')}e^{i\tilde{k}(\tilde{\theta}-\tilde{\theta}')}+\\
&e^{-ip(x-x')}\Big(e^{-i\tilde{k}\tilde{\theta}}\mathcal{G}^*(\tilde{k},\tilde{\theta}')+e^{i\tilde{k}\tilde{\theta}'}\mathcal{G}(\tilde{k},\tilde{\theta})+\mathcal{G}(\tilde{k},\tilde{\theta})\mathcal{G}^*(\tilde{k},\tilde{\theta}')\Big)-\\
&e^{ip(x-x')}\Big(e^{i\tilde{k}\tilde{\theta}}\mathcal{G}(\tilde{k},\tilde{\theta}')+e^{-i\tilde{k}\tilde{\theta}'}\mathcal{G}^*(\tilde{k},\tilde{\theta})+\mathcal{G}^*(\tilde{k},\tilde{\theta})\mathcal{G}(\tilde{k},\tilde{\theta}')\Big) \bigg).
\end{split}
\end{equation}
From this, the equal time commutation relation between DFRA field and its time derivative takes the form
\begin{equation}\label{dfr-LHS5}
\begin{split}
 \Big[\phi(x,\tilde{\theta}),\frac{d\phi(x',\tilde{\theta}')}{dt'}\Big]\Big|_{t=t'}=&i\delta^3(x-x')\int\frac{d^3\tilde{k}}{{(2\pi)^3}}~g(\tilde{k})~\bigg(\frac{e^{-i\tilde{k}(\tilde{\theta}-\tilde{\theta}')}}{2}+\frac{e^{i\tilde{k}(\tilde{\theta}-\tilde{\theta}')}}{2}+\frac{e^{-i\tilde{k}\tilde{\theta}}\mathcal{G}^*(\tilde{k},\tilde{\theta}')}{2}+\\
&\frac{e^{i\tilde{k}\tilde{\theta}'}\mathcal{G}(\tilde{k},\tilde{\theta})}{2}+\frac{e^{i\tilde{k}\tilde{\theta}}\mathcal{G}(\tilde{k},\tilde{\theta}')}{2}+\frac{e^{-i\tilde{k}\tilde{\theta}'}\mathcal{G}^*(\tilde{k},\tilde{\theta})}{2}+\frac{\mathcal{G}(\tilde{k},\tilde{\theta})\mathcal{G}^*(\tilde{k},\tilde{\theta}')}{2}+\\
&\frac{\mathcal{G}(\tilde{k},\tilde{\theta}')\mathcal{G}^*(\tilde{k},\tilde{\theta})}{2}\bigg)
\end{split}
\end{equation} 
If we demand that the equal time commutation relation between DFRA scalar field and its time derivative is undeformed, i.e, 
\begin{equation}\label{dfr-LHS6}
 \Big[\phi(x,\tilde{\theta}),\frac{d\phi(x',\tilde{\theta}')}{dt'}\Big]\Big|_{t=t'}=i\delta^3(x-x')\delta^3(\tilde{\theta}-\tilde{\theta}'),
\end{equation}  
a comparison of the RHS of Eq.(\ref{dfr-LHS5}) with the RHS of Eq.(\ref{dfr-LHS6}) leads to
\begin{equation}\label{dfr-LHS7}
\begin{split}
 \delta^3(\tilde{\theta}-\tilde{\theta}')=&\int\frac{d^3\tilde{k}}{{(2\pi)^3}}g(\tilde{k})\bigg(\frac{e^{-i\tilde{k}(\tilde{\theta}-\tilde{\theta}')}}{2}+\frac{e^{i\tilde{k}(\tilde{\theta}-\tilde{\theta}')}}{2}+\frac{e^{-i\tilde{k}\tilde{\theta}}\mathcal{G}^*(\tilde{k},\tilde{\theta}')}{2}+\frac{e^{i\tilde{k}\tilde{\theta}'}\mathcal{G}(\tilde{k},\tilde{\theta})}{2}+\\
&\frac{e^{i\tilde{k}\tilde{\theta}}\mathcal{G}(\tilde{k},\tilde{\theta}')}{2}+\frac{e^{-i\tilde{k}\tilde{\theta}'}\mathcal{G}^*(\tilde{k},\tilde{\theta})}{2}+\frac{\mathcal{G}(\tilde{k},\tilde{\theta})\mathcal{G}^*(\tilde{k},\tilde{\theta}')}{2}+\frac{\mathcal{G}(\tilde{k},\tilde{\theta}')\mathcal{G}^*(\tilde{k},\tilde{\theta})}{2}\bigg).
\end{split}
\end{equation}   
Using the definition for $\delta^3(\tilde{\theta}-\tilde{\theta}')$ on LHS of Eq.(\ref{dfr-LHS7}) and after some rearrangement, we get the explicit form of $g(\tilde{k})$ as
\begin{equation}\label{dfr-g}
 g(\tilde{k})=\frac{1}{1+
\frac{e^{i\tilde{k}\tilde{\theta}'}\mathcal{G}(\tilde{k},\tilde{\theta})+
e^{i\tilde{k}\tilde{\theta}}\mathcal{G}(\tilde{k},\tilde{\theta}')+
e^{-i\tilde{k}\tilde{\theta}'}\mathcal{G}^*(\tilde{k},\tilde{\theta})+
e^{-i\tilde{k}\tilde{\theta}}\mathcal{G}^*(\tilde{k},\tilde{\theta}')+
\mathcal{G}(\tilde{k},\tilde{\theta})\mathcal{G}^*(\tilde{k},\tilde{\theta}')+
\mathcal{G}(\tilde{k},\tilde{\theta}')\mathcal{G}^*(\tilde{k},\tilde{\theta})}{2\cos\tilde{k}(\tilde{\theta}-\tilde{\theta}')}}.
\end{equation}    
Therefore with the above choice of $g(\tilde{k})$ in Eq.(\ref{dfr-def}), the equal time commutation relation between DFRA scalar field and its time derivative becomes undeformed. Now for this particular choice of $g(\tilde{k})$, the deformed oscillator algebra in Eq.(\ref{dfr-def}) takes the form
\begin{equation}\label{dfr-def1}
 [a(p,\tilde{k}),a^{\dagger}(p',\tilde{k}')]=\frac{\delta^3(p-p')\delta^3(\tilde{k}-\tilde{k}')}{1+
\frac{e^{i\tilde{k}\tilde{\theta}'}\mathcal{G}(\tilde{k},\tilde{\theta})+
e^{i\tilde{k}\tilde{\theta}}\mathcal{G}(\tilde{k},\tilde{\theta}')+
e^{-i\tilde{k}\tilde{\theta}'}\mathcal{G}^*(\tilde{k},\tilde{\theta})+
e^{-i\tilde{k}\tilde{\theta}}\mathcal{G}^*(\tilde{k},\tilde{\theta}')+
\mathcal{G}(\tilde{k},\tilde{\theta})\mathcal{G}^*(\tilde{k},\tilde{\theta}')+
\mathcal{G}(\tilde{k},\tilde{\theta}')\mathcal{G}^*(\tilde{k},\tilde{\theta})}{2\cos\tilde{k}(\tilde{\theta}-\tilde{\theta}')}}.
\end{equation}
Here we see that the deformation factor in this oscillator algebra depends on the non-commutative coordinates $\tilde{\theta}$ as well as $\tilde{\theta'}$. This deformation factor is not unique as it changes with the choice of the weight function $W(\theta)$. Further this deformation factor reduces to unity when the weight function becomes one. It is to be noted that this $\theta$ dependent deformation is feature arising from non-commutativity. For the specific case $\theta=\theta'$, the deformation factor becomes, $g(\tilde{k})=\frac{1}{1+e^{-i\tilde{k}\tilde{\theta}}\mathcal{G}^*(\tilde{k},\tilde{\theta})+e^{i\tilde{k}\tilde{\theta}}\mathcal{G}(\tilde{k},\tilde{\theta})+\mathcal{G}(\tilde{k},\tilde{\theta})\mathcal{G}^*(\tilde{k},\tilde{\theta})}$. The deformed oscillator algebra valid up to first non-vanishing term in ${\theta}$, obtained from Eq.(\ref{dfr-def1}), is
\begin{equation}\label{dfr-def2}
 [a(p,\tilde{k}),a^{\dagger}(p',\tilde{k}')]=\delta^3(p-p')\delta^3(\tilde{k}-\tilde{k}')\bigg(1-\frac{(\tilde{k}\tilde{\theta'})^3}{12\tilde{k}^2\lambda^2}-\frac{(\tilde{k}\tilde{\theta})^3}{12\tilde{k}^2\lambda^2}\bigg).
\end{equation}
From Eq.(\ref{dfr-def2}), we observe that the first non-vanishing NC correction term in the deformed oscillator algebra depends on $\theta^3$. This is contrast with that in $\kappa$-deformed space-time, where the correction term is first order in the NC parameter \cite{vishnu1}. We observe that the correction terms depend on $1/\lambda^4$, where $\lambda$ is the length scale introduced by NC space-time. 

Note the presence of a weight function dependent term in the deformed equal time commutation relation between the DFRA field and its derivative in Eq.(\ref{dfr-LHS2}). These terms vanish when $W(\theta)=1$ and in this case, our results agree with that obtained in \cite{amorim2}. This weight function makes the action/Lagrangian non-unique and this non-uniqueness associated with the action/Lagrangian is a feature of the NC field theories resulting in a deformed oscillator algebra. Thus, in this section, we have derived the deformed commutation relation between the DFRA scalar field and its conjugate. For this, we showed that if this commutation relation is required to be undeformed, the corresponding quantised oscillator algebra becomes deformed.   


\section{Conserved currents}

In this section, we apply the procedure for constructing the conserved currents (from EOM alone) to the DFRA scalar theories. Using this scheme, we then derive the conserved currents corresponding to the translational as well as Lorentz symmetry for DFRA scalar field.
 
We construct the conserved currents from the EOM using the method discussed in sec.4.  For this we now reformulate the definition of $\Gamma_{\mu}(\partial,-\overleftarrow{\partial})$ operator compatible with the DFR space-time as $\Gamma_{A}(\partial,-\overleftarrow{\partial})$ and using Eq.(\ref{gamma}) we get
\begin{equation}\label{dfr-gamma}
 \begin{split}
 \Gamma_{A}(\partial,-\overleftarrow{\partial})&=\sum_{l=0}^{N-1}\sum_{i=0}^{l}\Lambda_{AA_1.....A_l}\partial_{A_1}.....\partial_{A_i}(-\overleftarrow{\partial}_{A_{i+1}})......(-\overleftarrow{\partial}_{A_l})\\
&=\Lambda_{A}+\Lambda_{AB}(\partial^{B}-\overleftarrow{\partial}^{B})+\Lambda_{ABC}(\partial^{B}\partial^{C}-\partial^{B}\overleftarrow{\partial}^{C}+\overleftarrow{\partial}^{B}\overleftarrow{\partial}^{C})+......
\end{split}
\end{equation}
Thus by substituting the relevant components of $\Lambda(\partial,\partial_{\theta})$, i.e, Eq.(\ref{dfr-lambda3}), in Eq.(\ref{dfr-gamma}) we obtain the $\Gamma_{A}(\partial,-\overleftarrow{\partial})$ explicitly as
\begin{equation}\label{dfr-gamma1}
 \Gamma_{A}(\partial,-\overleftarrow{\partial})=\Big(\partial_{\mu}-\overleftarrow{\partial}_{\mu},\lambda(\partial_{\theta_i}-\overleftarrow{\partial}_{\theta_i})-\frac{\epsilon\theta_i}{2\lambda^3}\Big).
\end{equation}
Using Eq.(\ref{current}), we get the generic conserved current corresponding to the scalar field in DFR space-time as 
\begin{equation}\label{dfr-current}
 J_A=\phi(x,\theta)\Gamma_A(\partial,-\overleftarrow{\partial})\delta\phi(x,\theta).
\end{equation}
Under translation symmetry, the coordinates in DFR space-time transform as
\begin{equation}\label{dfr-trans}
 x_{\mu}\to x_{\mu}+a_{\mu},~\theta_i\to\theta_i+b_i,
\end{equation}
where $a_{\mu}$ and $b_i$ are the parameters of translations in DFR space-time. Correspondingly the infinitesimal change in the field under this transformation is found to be
\begin{equation}\label{dfr-trans1}
 \delta\phi(x,\theta)=-a^{\mu}\partial_{\mu}\phi(x,\theta)-b^i\partial_{\theta_i}\phi(x,\theta)\equiv-C^B\partial_B\phi(x,\theta),
\end{equation}
where we denote $C^B=(a^{\mu},b^i/\lambda)$. Using these we write down the conserved currents associated with the translational symmetry in DFR space-time as
\begin{equation}\label{trans-curr}
 J_A=-C^B\phi(x,\theta)\Gamma_A(\partial,-\overleftarrow{\partial})\partial_B\phi(x,\theta).
\end{equation}
Now we get the explicit form of the components of the conserved currents corresponding to the translational symmetry in the DFR space-time by substituting Eq.(\ref{dfr-trans1}) and Eq.(\ref{dfr-gamma1}) in Eq.(\ref{trans-curr})
\begin{equation}\label{drf-trans2}
\begin{split}
 J_{\mu}=&a^{\nu}\partial_{\mu}\phi(x,\theta)\partial_{\nu}\phi(x,\theta)-a^{\nu}\phi(x,\theta)\partial_{\mu}\partial_{\nu}\phi(x,\theta)- b^i\phi(x,\theta)\partial_{\mu}\partial_{\theta_i}\phi(x,\theta)+b^i\partial_{\mu}\phi(x,\theta)\partial_{\theta_i}\phi(x,\theta),\\
 J_{\theta_j}=&a^{\nu}\partial_{\theta_j}\phi(x,\theta)\partial_{\nu}\phi(x,\theta)-a^{\nu}\phi(x,\theta)\partial_{\theta_j}\partial_{\nu}\phi(x,\theta)-b^i\phi(x,\theta)\partial_{\theta_j}\partial_{\theta_i}\phi(x,\theta)+b^i\partial_{\theta_j}\phi(x,\theta)\partial_{\theta_i}\phi(x,\theta)+\\
&a^{\nu}\frac{\theta_j}{2\lambda^3}\phi(x,\theta)\partial_{\nu}\phi(x,\theta)+b^{i}\frac{\theta_j}{2\lambda^3}\phi(x,\theta)\partial_{\theta_i}\phi(x,\theta).
\end{split}
\end{equation}
We notice from the above expressions that the Minkowskian part of the conserved current do not contain the weight function dependent terms (though $J_{\mu}$ does have $\theta$ dependent terms). But on the other hand, the $\theta$ components of conserved current contain two terms that depend on weight function. From this expression for the conserved current, we now obtain the explicit form of the energy-momentum tensor for the scalar field in DFR space-time as
\begin{equation}\label{tran-em}
 T_{AB}=\phi(x,\theta)\Gamma_A(\partial,-\overleftarrow{\partial})\partial_B\phi(x,\theta).
\end{equation}
Thus we write down the components of the energy momentum for a scalar field in DFR space-time as
\begin{equation}\label{dfr-em}
\begin{split}
 T_{\mu\nu}=&\phi(x,\theta)\partial_{\mu}\partial_{\nu}\phi(x,\theta)-\partial_{\mu}\phi(x,\theta)\partial_{\nu}\phi(x,\theta),\\
 T_{\mu\theta_i}=&\lambda\phi(x,\theta)\partial_{\mu}\partial_{\theta_i}\phi(x,\theta)-\lambda\partial_{\mu}\phi(x,\theta)\partial_{\theta_i}\phi(x,\theta),\\
 T_{\theta_i\mu}=&\lambda\phi(x,\theta)\partial_{\theta_i}\partial_{\mu}\phi(x,\theta)-\lambda\partial_{\theta_i}\phi(x,\theta)\partial_{\mu}\phi(x,\theta)-\frac{\theta_i}{2\lambda^3}\phi(x,\theta)\partial_{\mu}\phi(x,\theta),\\
 T_{\theta_i\theta_j}=&\lambda^2\phi(x,\theta)\partial_{\theta_i}\partial_{\theta_j}\phi(x,\theta)-\lambda^2\partial_{\theta_i}\phi(x,\theta)\partial_{\theta_j}\phi(x,\theta)-\frac{\theta_i}{2\lambda^2}\phi(x,\theta)\partial_{\theta_j}\phi(x,\theta).
\end{split}
\end{equation}
From above, we observe that the last three components of the energy-momentum tensor are purely due to the presence of $\theta$ coordinate in DFR space-time. Here components of $T_{\mu\nu}$ is symmetric, as in the commutative case. But $T_{\mu\theta_i}\neq T_{\theta_i\mu}$ and $T_{\theta_i\theta_j}$ is not symmetric. It is quite interesting to see that $T_{\mu\theta_i}$ and $T_{\theta_i\mu}$ components become equal when the weight function reduces to unity. In the same way $T_{\theta_i\theta_j}$ also becomes symmetric when the weight function becomes unity. Thus we see that the energy-momentum tensor of the DFRA scalar field ceases to be symmetric in its indices due to the weight function $W(\theta)$.  

Note that one can also obtain the expression for the momenta corresponding to the scalar field in DFR space-time using the definition for $T_{AB}$ as 
\begin{equation}\label{dfr-moment}
 P_B=\int d^3x~d^3\theta~W(\theta)T_{0B}.
\end{equation}
Now we analyse the Lorentz symmetry in DFR space-time. Under the Lorentz transformation the DFR space-time coordinates transform as
\begin{equation}\label{dfr-lorentz}
 x_{\mu}\to x_{\mu}+\omega_{\mu}^{~\nu}x_{\nu},~\theta_i\to\theta_i+\omega_i^{~j}\theta_j.
\end{equation}
Using the above transformation rule, we see that under the Lorentz transformation the DFRA scalar field changes as 
\begin{equation}\label{lorentz-field}
\begin{split}
 \delta\phi(x,\theta)=&-\omega^{\mu\nu}x_{\nu}\partial_{\mu}\phi(x,\theta)-\omega^{ij}\theta_j\partial_{\theta_i}\phi(x,\theta)\equiv-C^{AB}\phi(x,\theta)X_B\partial_A\phi(x,\theta),
\end{split}
\end{equation}
where we denote $C^{AB}=((\omega^{\mu\nu},0),(0,\omega^{ij}))$ and $X_B=(x_{\mu},{\theta_i}/{\lambda})$.

Thus the expression for the conserved current corresponding to this transformation of the scalar field in DFR space-time can be written as
\begin{equation}\label{lorentz-ang}
 J_A=-C^{CB}\phi(x,\theta)\Gamma_A(\partial,-\overleftarrow{\partial})X_B\partial_C\phi(x,\theta)\equiv C^{BC}M_{ABC},
\end{equation}
where,
\begin{equation}\label{mabc}
 M_{ABC}=\phi(x,\theta)\Gamma_A(\partial,-\overleftarrow{\partial})X_B\partial_C\phi(x,\theta).
\end{equation}
Substituting Eq.(\ref{lorentz-field}) and Eq.(\ref{dfr-gamma1}) in Eq.(\ref{lorentz-ang}), we obtain the expression for the components of the conserved currents corresponding to Lorentz symmetry as
\begin{equation}\label{lorentz-current}
\begin{split}
 J_{\mu}=&\omega_{\mu}^{~\nu}\phi(x,\theta)\partial_{\nu}\phi(x,\theta)+\omega^{\nu\lambda}\phi(x,\theta)x_{\nu}\partial_{\lambda}\phi(x,\theta)-\omega^{\nu\lambda}\partial_{\mu}\phi(x,\theta)x_{\nu}\partial_{\lambda}\phi(x,\theta)+\\
&\omega^{ij}\phi(x,\theta)\theta_i\partial_{\mu}\partial_{\theta_j}\phi(x,\theta)-\omega^{ij}\partial_{\mu}\theta_i\partial_{\theta_j}\phi(x,\theta),\\
J_{\theta_i}=& \lambda\omega^{\mu\nu}\phi(x,\theta)x_{\mu}\partial_{\theta_i}\partial_{\nu}\phi(x,\theta)-\lambda\omega^{\mu\nu}\partial_{\theta_i}\phi(x,\theta)x_{\mu}\partial_{\nu}\phi(x,\theta)+\lambda\omega_{i}^{~k}\phi(x,\theta)\partial_{\theta_k}\phi(x,\theta)+\\
&\lambda\omega^{jk}\phi(x,\theta)\theta_j\partial_{\theta_i}\partial_{\theta_k}\phi(x,\theta)-\frac{\theta_i}{2\lambda^3}\omega^{\mu\nu}\phi(x,\theta)x_{\mu}\partial_{\nu}\phi(x,\theta)-\frac{\theta_i}{2\lambda^3}\omega^{jk}\phi(x,\theta)\theta_{j}\partial_{\theta_k}\phi(x,\theta).
\end{split}
\end{equation}
As in the case of translational symmetry, here also we see that only the $\theta$ components of the conserved current pick up weight function dependent terms. 

Using the defintion for $M_{ABC}$ given in Eq.(\ref{mabc}), we write down the expression for Lorentz generator as,
\begin{equation}
\begin{split}
 M_{BC}=&\int d^3x~d^3\theta~W(\theta) M_{0BC}\\
 =&\int d^3x~d^3\theta~W(\theta)\Big(\phi(x,\theta)\Gamma_0(\partial,-\overleftarrow{\partial})X_B\partial_C\phi(x,\theta)\Big).
\end{split}
\end{equation}  
The explicit form of the components of this tensor corresponding to the scalar field in DFR space-time are
\begin{equation}\label{lorentz-ang1}
\begin{split}
 M_{\mu\nu}=&\int d^3x~d^3\theta~W(\theta)\bigg(\phi(x,\theta)\delta_{\mu 0}\partial_{\nu}\phi(x,\theta)+\phi(x,\theta)x_{\mu}\partial_0\phi(x,\theta)\partial_{\nu}\phi(x,\theta)-\partial_0x_{\mu}\partial_{\nu}\phi(x,\theta)-\\
&\phi(x,\theta)\delta_{\nu 0}\partial_{\mu}\phi(x,\theta)-\phi(x,\theta)x_{\nu}\partial_0\partial_{\mu}\phi(x,\theta)+\partial_0\phi(x,\theta)x_{\nu}\partial_{\mu}\phi(x,\theta)\bigg),\\
M_{\mu\theta_i}=&\int d^3x~d^3\theta~W(\theta)\bigg(\lambda\phi(x,\theta)\delta_{\mu 0}\partial_{\theta_i}\phi(x,\theta)+\lambda\phi(x,\theta)x_{\mu}\partial_0\partial_{\theta_i}\phi(x,\theta)-\lambda\partial_0\phi(x,\theta)x_{\mu}\partial_{\theta_i}\phi(x,\theta)-\\
&\frac{\theta_i}{\lambda}\phi(x,\theta)\partial_0\partial_{\mu}\phi(x,\theta)+\frac{\theta_i}{\lambda}\partial_0\phi(x,\theta)\partial_{\mu}\phi(x,\theta)\bigg),\\
M_{\theta_i\mu}=&\int d^3x~d^3\theta~W(\theta)\bigg(-\lambda\phi(x,\theta)\delta_{\mu 0}\partial_{\theta_i}\phi(x,\theta)-\lambda\phi(x,\theta)x_{\mu}\partial_0\partial_{\theta_i}\phi(x,\theta)+\lambda\partial_0\phi(x,\theta)x_{\mu}\partial_{\theta_i}\phi(x,\theta)+\\
&\frac{\theta_i}{\lambda}\phi(x,\theta)\partial_0\partial_{\mu}\phi(x,\theta)-\frac{\theta_i}{\lambda}\partial_0\phi(x,\theta)\partial_{\mu}\phi(x,\theta)\bigg),\\
M_{\theta_i\theta_j}&=\int d^3x~d^3\theta~W(\theta)\bigg(\phi(x,\theta)\theta_i\partial_0\partial_{\theta_j}\phi(x,\theta)-\partial_0\phi(x,\theta)\theta_i\partial_{\theta_j}\phi(x,\theta)-\phi(x,\theta)\theta_j\partial_0\partial_{\theta_i}\phi(x,\theta)+\\
&\partial_0\theta_j\partial_{\theta_i}\phi(x,\theta)\bigg).
\end{split}
\end{equation}
From the above, we observe that all the components of the generators of Lorentz transformation of scalar field in DFR space-time depend on weight function through an overall multiplication factor $W(\theta)$ coming from the measure. In the $\lim\lambda\to 0$ with $\theta=0$, the components like $M_{\mu\theta_i}$, $M_{\theta_i\nu}$, $M_{\theta_i\theta_j}$ vanishes and $M_{\mu\nu}$ becomes that of the commutative scalar field. 

\section{Response of detector coupled to DFRA scalar field and Unruh effect} 


In this section, we study how quantum gravity effects brought in by the non-commutativity of this space-time affects the vacuum state by studying the Unruh effect in DFR space-time. Here we study the response of a massless DFRA scalar field that is coupled to Unruh-DeWitt detector \cite{unruh,birrel} moving along a uniformly accelerated trajectory. For this, we first obtain the expression for the positive Wightman function, and using this, we then derive the explicit form of the response function in both even and odd dimensions of the DFR space-time. 

Let us now choose a detector of monopole moment $m(\tau)$ whose path is parametrised by the proper time $\tau$ and it is coupled to a massless DFRA scalar field. We describe this interaction term as in the commutative space-time. Therefore the interacting Lagrangian is taken as
\begin{equation}\label{interaction}
 \mathcal{L}_{int}=m(\tau)\phi\big(x(\tau),\tilde{\theta}(\tau)\big).
\end{equation} 
We consider a situation where the field is in vacuum state $\ket{0}$ and the detector is in its ground state $\ket{E_0}$. This detector will make a transition to an excited state $\ket{E}$, where $E>E_0$, as it moves along a trajectory with constant acceleration. This will affect the massless DFRA field coupled to the detector and the field makes a transition to an excited state $\ket{\psi}$. Using the time-independent first order perturbation theory we find the expression for the transition amplitude as
\begin{equation}\label{ampli}
 \mathcal{M}_{i\to f}=i\bra{E,\psi}\int_{-\infty}^{\tau_0}d\tau~m(\tau)\phi\big(x(\tau),\tilde{\theta}(\tau)\big)\ket{0,E_0}.
\end{equation}
Under the time evolution the detectors monopole moment evolve as $m(\tau)=e^{iH_0\tau}m(0)e^{-iH_0\tau}$. Here $H_0$ corresponds to the Hamiltonian of the detector and we have $H_0\ket{E}=E\ket{E}$ as well as $H_0\ket{E_0}=E_0\ket{E_0}$. Thus from the above Eq.(\ref{ampli}), the transition probability can be written as 
\begin{equation}\label{trans}
 \big|\mathcal{M}_{i\to f}\big|^2=\sum_E\Big|\bra{E}m(0)\ket{E_0}\Big|^2\mathcal{F}(\Delta E)
\end{equation}
where $\mathcal{F}(\Delta E)$ is the response function, whose explicit form is
\begin{equation}\label{response}
 \mathcal{F}(\Delta E)=\int_{-\infty}^{\tau_0}\int_{-\infty}^{\tau_0}d\tau d\tau'e^{-i\Delta E(\tau-\tau')}G^+\big(x(\tau),\tilde{\theta}(\tau);x(\tau'),\tilde{\theta}(\tau')\big).
\end{equation}
and $\Delta E=E-E_0$. In above expression, $G^+\big(x(\tau),\tilde{\theta}(\tau);x(\tau'),\tilde{\theta}(\tau')\big)$ is the positive Wightman function and it is defined as
\begin{equation}\label{wightman1}
\begin{split}
 G^+(x,\tilde{\theta};x',\tilde{\theta}')=\bra{0}\phi(x,\tilde{\theta})\phi(x',\tilde{\theta}')\ket{0}.
\end{split}
\end{equation}  
It is straightforward to obtain the Wightman function from the EOM alone. But in this work, we calculate the Wightman function from the vacuum expectation value of the quantised DFRA scalar fields. We adopt this procedure to see the effects of the quantised DFRA fields on the response function. Moreover, this method also helps to analyse the effects of undeformed as well as deformed oscillator algebra on the Wightman function. 

We obtain the expression for the positive Wightman function in $(1+3+d_{\theta})$ dimension \footnote{Here $d_{\theta}$ is the number of dimension introduced by NC in DFR space-time} DFR space-time by taking the vacuum expectation value of the DFRA field operators given in Eq.(\ref{phi1}). Using the explicit form of $\mathcal{G}(\tilde{k},\tilde{\theta})$ from Eq.(\ref{gk}) in Eq.(\ref{wightman1}), we calculate the Wightman function valid up to first order in $1/\lambda^2$ (note that in this calculation we assume that the DFRA oscillator algebra is undeformed) and we get, 
\begin{equation}\label{wightman3}
\begin{split}
 G^+(x,\tilde{\theta};x',\tilde{\theta}')=&\frac{N(d_{\theta})}{\Big[(x-x')^2+(\tilde{\theta}-\tilde{\theta'})^2-(t-t')^2\Big]^{(2+d_{\theta})/2}}-\frac{(\tilde{\theta}')^4}{4\lambda^2}\frac{N(d_{\theta})(2+d_{\theta})}{\Big[(x-x')^2+\tilde{\theta}^2-(t-t')^2\Big]^{(4+d_{\theta})/2}}\\
&-\frac{(\tilde{\theta})^4}{4\lambda^2}\frac{(2+d_{\theta})}{\Big[(x-x')^2+\tilde{\theta'}^2-(t-t')^2\Big]^{(4+d_{\theta})/2}}+\frac{i a_1(\tilde{\theta})^4}{6\lambda^2}\frac{N(d_{\theta})(2+d_{\theta})}{\Big[(x-x')^2+\tilde{\theta'}^2-(t-t')^2\Big]^{(4+d_{\theta})/2}}\\
&-\frac{i a_1(\tilde{\theta}')^4}{6\lambda^2}\frac{N(d_{\theta})(2+d_{\theta})}{\Big[(x-x')^2+\tilde{\theta}^2-(t-t')^2\Big]^{(4+d_{\theta})/2}}-
\frac{i a_1(\tilde{\theta})^5}{12\lambda^2}\frac{N(d_{\theta})(2+d_{\theta})(4+d_{\theta})}{\Big[(x-x')^2+\tilde{\theta'}^2-(t-t')^2\Big]^{(6+d_{\theta})/2}}\\
&+\frac{i a_1(\tilde{\theta})^5}{12\lambda^2}\frac{N(d_{\theta})(2+d_{\theta})(4+d_{\theta})}{\Big[(x-x')^2+\tilde{\theta'}^2-(t-t')^2\Big]^{(6+d_{\theta})/2}}
\end{split}
\end{equation}
where, $N(d_{\theta})=\frac{\Gamma\big(\frac{2+d_{\theta}}{2}\big)}{4\pi^{(4+d_{\theta})/2}}$. The origin of all $1/\lambda^2$ dependent terms in the above can be traced back to $\mathcal{G}(\tilde{k},\tilde{\theta}),~\mathcal{G}^*(\tilde{k},\tilde{\theta})$ dependent terms in Eq.(\ref{phi1}). Since when $W(\theta)=1$, $\mathcal{G}(\tilde{k},\tilde{\theta})=\mathcal{G}^*(\tilde{k},\tilde{\theta})=0$, we see that all these terms drop out from the above equation. The last six terms are not unique as they change when we choose a different weight function. Note that here we obtain the positive Wightman function in the commutative $\lim \theta\to 0$ and $d_{\theta}\to 0$.

Now we assume that the detector is moving along a trajectory with a uniform acceleration. The coordinates in this uniformly accelerated frame is defined as
\begin{equation}\label{uniaccl}
 t(\tau)=\frac{1}{A}\sinh A\tau,~x(\tau)=\frac{1}{A}\cosh A\tau,~y=\textnormal{constant},~z=\textnormal{constant},~\tilde{\theta}_i=\textnormal{constant (say } \theta),
\end{equation}
where $A$ is the constant proper acceleration. In the above we have choosen the detector to be accelerating within the commutative coordinates and we have fixed the extra NC coordinates to be a constant. Thus even in the commutative limit, i.e, $\theta\to 0$, the detector will continue to move in a uniformly accelerating trajectory. Now in this uniformly accelerating path given in Eq.(\ref{uniaccl}), the positive Wightman function takes the form
\begin{equation}\label{wightman2}
\begin{split}
 G^+(\tau-\tau')=&\Big(\frac{A}{2}\Big)^{2+d_{\theta}}\frac{N(d_{\theta})}{\Big(\sinh^2\frac{A(\tau-\tau')}{2}\Big)^{(2+d_{\theta})/2}}
-\frac{\theta^4}{2\lambda^2}\Big(\frac{A}{2}\Big)^{4+d_{\theta}}\frac{N(d_{\theta})(2+d_{\theta})}{\Big(\sinh^2\frac{A(\tau-\tau')}{2}+\frac{A^2\theta^2}{4}\Big)^{(4+d_{\theta})/2}}.
\end{split}
\end{equation}
From the above equation, we observe that the positive Wightman function in the uniformly accelerating coordinate possess $\theta$ dependent correction terms and as mentioned above, these terms are not unique, and it changes with the choice of weight function. From Eq.(\ref{wightman2}), we obtain the commutative case in the limit $\theta=0$ and $d_{\theta}\to 0$.
 
The rate of transition probability can be obtained from Eq.(\ref{trans}) as
\begin{equation}\label{rate}
 \mathcal{T}(\Delta E)=\frac{d\big|\mathcal{M}_{i\to f}\big|^2}{d\tau_0}=\sum_E\Big|\bra{E}m(0)\ket{E_0}\Big|^2\frac{d\mathcal{F}(\Delta E)}{d\tau_0}.
\end{equation}
By evaluating $\frac{d\mathcal{F}(\Delta E)}{d\tau_0}$ using the explicit form of the response function given in Eq.(\ref{response}) and simplifying it, we get the expression for the transition probability rate as
\begin{equation}\label{rate1}
 \mathcal{T}(\Delta E)=\sum_E\Big|\bra{E}m(0)\ket{E_0}\Big|^2\int_{-\infty}^{\infty}d\tau~e^{-i\Delta E\tau}G^+(\tau).
\end{equation}
Using the expression for the positive energy Wightman function, we will now calculate the transition probability rate in DFR space-time. Substituting Eq.(\ref{wightman2}) in Eq.(\ref{rate1}) we get 
\begin{equation}\label{rate2}
\begin{split}
 \mathcal{T}(\Delta E)=&\sum_E\Big|\bra{E}m(0)\ket{E_0}\Big|^2\Big(\frac{A}{2}\Big)^{2+d_{\theta}}N(d_{\theta})\int_{-\infty}^{\infty}d\tau~\frac{e^{-i\Delta E\tau}}{\Big(\sinh^2\frac{A\tau}{2}\Big)^{(2+d_{\theta})/2}}-\\
&\sum_E\Big|\bra{E}m(0)\ket{E_0}\Big|^2\frac{\theta^4}{2\lambda^2}\Big(\frac{A}{2}\Big)^{4+d_{\theta}}N(d_{\theta})(2+d_{\theta})\int_{-\infty}^{\infty}d\tau~\frac{e^{-i\Delta E\tau}}{\Big(\sinh^2\frac{A\tau}{2}+\frac{A^2\theta^2}{4}\Big)^{(4+d_{\theta})/2}}.
\end{split}
\end{equation}
We solve these integrals \cite{grad} and after further simplifications, we get the explicit form of the transition probability rate in $(1+3+d_{\theta})$ dimensional DFR space-time as
\begin{equation}\label{rate3}
\begin{split}
 \mathcal{T}(\Delta E)=&\sum_E\Big|\bra{E}m(0)\ket{E_0}\Big|^2\bigg(\xi_0^{(d_{\theta})}-\frac{(A\theta)^4}{32\lambda^2}\xi_1^{(d_{\theta})} \bigg)\frac{1}{e^{2\pi\Delta E/A}+(-1)^{d_{\theta}+1}},
\end{split}
\end{equation} 
where,
\begin{equation}\label{xi0}
 \xi_0^{(d_{\theta})}(\Delta E,A)=\frac{\Gamma\big(\frac{d_{\theta}+2}{2}\big)A^{(1+d_{\theta})}}{2\pi^{(2+d_{\theta})/2}}
 \begin{cases}
   \displaystyle{\prod_{k=0}^{d_{\theta}/2}}\bigg(k^2+\frac{(\Delta E)^2}{A^2}\bigg)\frac{A}{\Delta E},~~~~~~~~~~~~~~\textnormal{if $d_{\theta}$ is even}\\\\
   \displaystyle{\prod_{k=0}^{(d_{\theta}-1)/2}}\bigg(\frac{(2k+1)^2}{4}+\frac{(\Delta E)^2}{A^2}\bigg),~~~~~\textnormal{if $d_{\theta}$ is odd}
 \end{cases}
\end{equation}
and
\begin{equation}\label{xi1}
\begin{split}
 \xi_1^{(d_{\theta})}(\Delta E,A,\theta)=&\frac{2^{(4+d_{\theta})}\pi e^{\pi\Delta E/A}}{\Gamma(4+d_{\theta})A}\frac{\bigg(\frac{A^2\theta^2}{2}-1+A\theta\sqrt{\frac{A^2\theta^2}{4}-1}\bigg)^{i\Delta E/A}}{\bigg(\frac{A^2\theta^2}{2}-1-A\theta\sqrt{\frac{A^2\theta^2}{4}-1}\bigg)^{(4+d_{\theta})/2}}\times\\
&{}_2F_1\Bigg(\frac{4+d_{\theta}}{2};\frac{i\Delta E}{A}+\frac{4+d_{\theta}}{2};4+d_{\theta};
1-\frac{e^{(4+d_{\theta})/2}}{\Big(\frac{A^2\theta^2}{2}-1+A\theta\sqrt{\frac{A^2\theta^2}{4}-1}\Big)}\Bigg)\times\\
 &\begin{cases}
   \displaystyle{\prod_{k=1}^{d_{\theta}}}\bigg(k^2+\frac{(\Delta E)^2}{A^2}\bigg),~~~~~~~~~~~~~~\textnormal{if $d_{\theta}$ is even}\\\\
   \displaystyle{\prod_{k=1}^{d_{\theta}}}\bigg(\frac{(2k-1)^2}{4}+\frac{(\Delta E)^2}{A^2}\bigg),~~~~~\textnormal{if $d_{\theta}$ is odd}
 \end{cases}
\end{split}
\end{equation}
In Eq.(\ref{rate3}), we notice that the transition probability rate contains $\theta$ dependent term and they also possess the same distribution function as that in the commutative case. 

Now we get the expression for the transition probability rate with $d_{\theta}=1$ as
\begin{equation}\label{d=1}
\begin{split}
 \mathcal{T}(\Delta E)=&\sum_E\Big|\bra{E}m(0)\ket{E_0}\Big|^2\frac{A^2}{4\pi}\Big(\frac{1}{4}+\frac{(\Delta E)^2}{A^2}\Big)\Bigg[1
-\frac{A\theta^4\pi^2e^{\pi\Delta E/A}}{6\lambda^2}\frac{\bigg(\frac{A^2\theta^2}{2}-1+A\theta\sqrt{\frac{A^2\theta^2}{4}-1}\bigg)^{i\Delta E/A}}{\bigg(\frac{A^2\theta^2}{2}-1-A\theta\sqrt{\frac{A^2\theta^2}{4}-1}\bigg)^{5/2}}\times\\
&{}_2F_1\Bigg(\frac{5}{2};\frac{i\Delta E}{A}+\frac{5}{2};5;
1-\frac{e^{5/2}}{\Big(\frac{A^2\theta^2}{2}-1+A\theta\sqrt{\frac{A^2\theta^2}{4}-1}\Big)}\Bigg)\Bigg]\frac{1}{e^{2\pi\Delta E/A}+1}.
\end{split}
\end{equation}
From above we observe that for $d_{\theta}=1$ (i.e, (1+3)+1=5, odd-dimensional DFR space-time), $\mathcal{T}(\Delta E)$ picks up a Fermi-Dirac (FD) distribution factor. Similarly we obtain the expression for the transition probability rate with $d_{\theta}=2$ as
\begin{equation}\label{d=2}
\begin{split}
 \mathcal{T}(\Delta E)=&\sum_E\Big|\bra{E}m(0)\ket{E_0}\Big|^2\frac{A^2}{2\pi^2}\Big(1+\frac{(\Delta E)^2}{A^2}\Big)\Bigg[\Delta E
-\frac{A\theta^4\pi^2e^{\pi\Delta E/A}}{30\lambda^2}
\frac{\bigg(\frac{A^2\theta^2}{2}-1+A\theta\sqrt{\frac{A^2\theta^2}{4}-1}\bigg)^{i\Delta E/A}}{\bigg(\frac{A^2\theta^2}{2}-1-A\theta\sqrt{\frac{A^2\theta^2}{4}-1}\bigg)^{3}}\times\\
&\Big(4+\frac{(\Delta E)^2}{A^2}\Big)
{}_2F_1\Bigg(3;\frac{i\Delta E}{A}+3;6;1-\frac{e^{3}}{\Big(\frac{A^2\theta^2}{2}-1+A\theta\sqrt{\frac{A^2\theta^2}{4}-1}\Big)}\Bigg)\Bigg]\frac{1}{e^{2\pi\Delta E/A}-1}.
\end{split}
\end{equation}
Here we see that when $d_{\theta}=2$ (i.e, (1+3)+2=6, even dimensional DFR space-time), then $\mathcal{T}(\Delta E)$ picks up a Bose-Einstein (BE) distribution factor. Therefore we conclude that for an even dimensional DFR space, the transition probability rate has BE distribution, and for odd-dimensional DFR space, it has a FD distribution. Hence this result is in agreement with that in the commutative case \cite{takagi,crispino}.  

Here we also point out that the non-commutative correction terms in $\mathcal{T}(\Delta E)$ is not unique and it depends on the choice of the weight function. Note that from Eq.(\ref{rate3}), we get the commutative limit by taking $\theta=0$ and $d_{\theta}\to 0$.   

BE or FD distribution gets multiplied by an overall factor that depends on the NC of space-time (see Eq.(\ref{rate3})) and the acceleration $A$ of the detector in the DFRA framework, and thus, the temperature (Unruh temperature) associated with the transition probability rate is exactly the same as that in the commutative regime, $T=\frac{2\pi}{A}$. Similar results have been observed in $\kappa$-Minkowski space-time also \cite{kim,ravi}. Unlike the $\kappa$ Minkowski space-time, here the response function is independent of proper time $\tau_0$ and this is due to the fact that DFR space-time framework preserves the Lorentz symmetry.

\section{Conclusion}

We have studied the quantisation of the DFRA scalar field theory from its EOM, without referring to its Lagrangian. The DFRA scalar field action is not unique due to the presence of the $\theta$ dependent weight function that has been introduced to avoid the divergences in the perturbative QFT. Thus the Lagrangian associated with the NC field theories are not unique and therefore, it is appropriate to use the EOM of the DFRA scalar field for quantising it. Also, note that this EOM is different from the EOM coming from Casimir of DFRA algebra as the Casimir is independent of the weight function. But in the appropriate limit former reduces to the latter.

By generalising the quantisation scheme, summarised in sec.4, to DFR space-time, we obtained the deformed equal time commutation relation between the DFRA field and its derivative, valid to all orders in the non-commutative parameter, when the DFRA oscillator algebra has the same form as that in the commutative case. It is shown that the first non-vanishing correction of this deformed algebra scales as $\frac{1}{\lambda^4}$ where $\lambda$ is the non-commutative length scale. Consistency of the condition that the commutation relation between the DFRA scalar field and its adjoint is not modified by NC is shown to lead to a deformed oscillator algebra. We observe that the first non-vanishing correction terms in the DFRA oscillator algebra depends on $\theta^3$ and $1/\lambda^4$. The commutation relation between the DFRA field and its adjoint obtained in \cite{amorim3} is not deformed as the weight function is chosen to be one, and this is in agreement with our result in the limit $W(\theta)=1$. In all these cases, we find that these deformation depends on the choice of the weight function. Deformed algebra of oscillators have also been obtained in \cite{vishnu1,vishnu2} while quantising the Klein-Gordon and Dirac fields in $\kappa$-Minkowski space-time, using the above procedure. Thus one sees that the deformed oscillator is a general feature of the NC field theories.

The conserved currents corresponding to translational as well as Lorentz symmetries for the DFRA scalar field have been derived from their EOM. This enables us to calculate the energy-momentum tensor and angular momentum of the DFRA scalar theory. We find that $T_{\theta_i\mu}$ components are not symmetric and this asymmetry is accounted due to the weight function. Such non-symmetric energy-momentum tensor for the scalar field has been obtained in Moyal space-time also \cite{abou}. 

Unruh effect for DFRA scalar field has been analysed by calculating the transition probability rate of uniformly accelerating detector, coupled to massless DFRA scalar field, between vacuum state and an excited state. We observe that this transition probability get an overall multiplication factor which depends on the non-commutative parameter. The thermal distribution part of this transition probability also gets modified, which depends on the dimension of the extra non-commutative directions (i.e., $\theta$-directions) introduced in the DFR space-time. We find that the transition probability becomes either Bose-Einstein or Fermi-Dirac distribution depending on whether the number of the extra dimensions introduced in the DFR space-time is even or odd. Importantly, we find that Unruh temperature is not modified.  

In this work, we have modelled the monopole detector by the commutative itself. Instead, one can also realise a NC version for this monopole moment, depending on the $1/\lambda^2$ terms. This would then give an additional $1/\lambda^2$ term to the Wightman function and transition probability. It will be interesting to analyse the Unruh effect in the DFR space-time using the Bogoliubov transformations. Such studies were done in both Moyal \cite{sachin} and $\kappa$-deformed \cite{vishnu1} space-times and we plan to take up this in future.

\subsection*{\bf Acknowledgments}
VR thanks Govt. of India, for support through DST-INSPIRE/IF170622.

\end{document}